\documentclass[twocolumn,showpacs,preprintnumbers,amsmath,amssymb,showkeys]{revtex4}

\usepackage{graphicx}
\usepackage{dcolumn}
\usepackage{bm}
\usepackage{multirow}

\begin{document}


\title{Critical Behavior of the $\boldsymbol{q}$ = 3,\,4-Potts model on Quasiperiodic Decagonal Lattices} 

\author{Carlos Handrey Araujo Ferraz}
\email{handrey@ufersa.edu.br}
\affiliation{Department of Exact and Natural Sciences, Universidade Federal Rural do Semi-\'Arido-UFERSA, PO Box 0137, CEP 59625-900, Mossor\'o, RN, Brazil}


\date{May 27, 2015}
\begin{abstract}
In this study, we performed Monte Carlo simulations of the $q=3,4$-Potts model on quasiperiodic decagonal lattices (QDL) to assess the critical behavior of these systems. Using the single histogram technique in conjunction with the finite-size scaling analysis, we estimate the infinite lattice critical temperatures and the leading critical exponents for $q=3$ and $q=4$ states.  Our estimates for the critical exponents on QDL are in good agreement with the exact values on 2D periodic lattices, supporting the claim that both the $q=3$ and $q=4$ Potts model on quasiperiodic lattices belong to the same universality class as those on 2D periodic lattices.
\end{abstract}

\pacs{61.44.Br; 75.10.Hk; 05.10.Ln } 
\keywords{Quasiperiodic decagonal lattice; $q$-Potts model; Critical exponents; Monte Carlo simulation} 
\maketitle

\section{Introduction \label{sec:int}}
Electron diffraction patterns exhibiting octagonal, decagonal, dodecagonal, and icosahedral point symmetry are found in various alloys. The most
well-known pattern is the icosahedral phase in $Al$-$Mn$ alloys, which is observed when these materials are cooled at a rapid rate such that their
constituent atoms do not have adequate time to form a crystal lattice. These structures are referred to as quasicrystals~\cite{shechtman84,levine86}. In principle, quasicrystals are characterized as atomic structures that present long-range quasiperiodic translational and long-range orientational order. They can exhibit rotational symmetries otherwise forbidden to crystals. In the last decade, quasicrystals have attracted significant attention, mostly because of their stronger magnetic properties and enhanced elasticity at higher temperatures, compared with the traditional crystals.

A most intriguing research topic about quasicrystals is to determine whether its intrinsic complicated structure can result in a change of the universality class compared with its counterpart periodic structure.  To this end, Potts model~\cite{potts52} offers a simple and feasible way to study quasicrystals from this perspective, as it contains both first- and second-order phase transitions.  However, given the lack of periodicity of these quasiperiodic lattices, only numerical approaches can be performed. Previous Monte Carlo studies on the ferromagnetic Potts model for quasiperiodic lattices~\cite{wilson88,wilson89,ledue97,xiong99,fu2006,bin2011} have revealed that both the systems belong to the same universality class, despite the critical temperature of the quasiperiodic lattices being higher than that of the square lattices. However, given the great variety of existing quasiperiodic lattices, this query has not been solved completely. Consequently, this necessitates extensive computational research for accurately estimating the static critical exponents in these lattices. To the best of our knowledge, studies concerning the $q=4$ Potts model on quasiperiodic lattices have been rarely reported in the literature.

\begin{figure}[!t]
 \centering
 \includegraphics*[scale=0.40,angle=0]{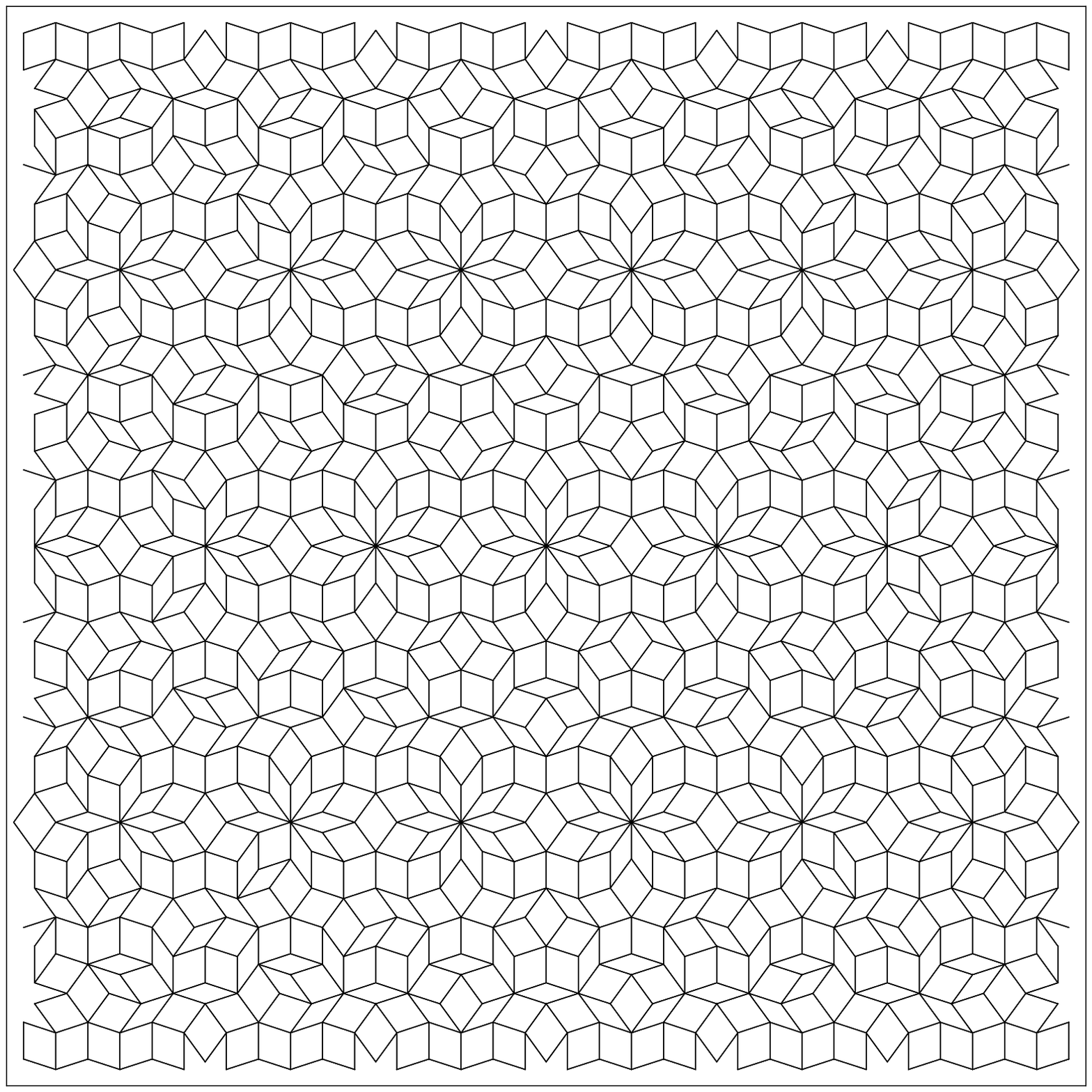}
 \caption{\label{fig:1}A periodic approximation of the QDL generates by the strip projection method. The lattice is shown inside a square projection window. The periodic boundary conditions are imposed at lattice sites closer to the projection window.}
\end{figure}

The present study investigates the critical behavior of the ferromagnetic $q= 3,4$-Potts model on quasiperiodic decagonal lattices (QDL) to accurately estimate the infinite QDL critical temperature and critical exponents for each case. An interesting example of a natural structure which presents a decagonal symmetry is the $Al_{71}Ni_{24}Fe_{5}$ quasicrystal found in the Khatyrka meteorite~\cite{luca}. The quasiperiodic lattices analyzed in this study were generated using the strip projection method~\cite{duneau85,conway86,vogg96} with spins placed in the vertices of the rhombi that constitute the QDL (Fig.~\ref{fig:1}). Periodic boundary conditions were applied on these lattices to avoid the boundary effects caused by the finite size.

This paper is organized as follows. Section \ref{sec:proc} briefly describes the strip projection method adopted for generating the QDL and periodic boundary conditions used in the simulations. Details of the Potts model and Monte Carlo simulation approach are described in section \ref{sec:mms}.  In section \ref{sec:f}, a succinct description of the finite-size scaling (FSS) relations used in the study is presented.  In section \ref{sec:r}, we present the results for $q=3$ and $q=4$ Potts model and compare them with previous results on quasi-periodic lattices. In section \ref{sec:c}, we conclude by summarizing the results and providing recommendations for further research.

\begin{figure*}[t]
\begin{minipage}[!r]{0.49\linewidth}
\includegraphics*[scale=0.30,angle=0]{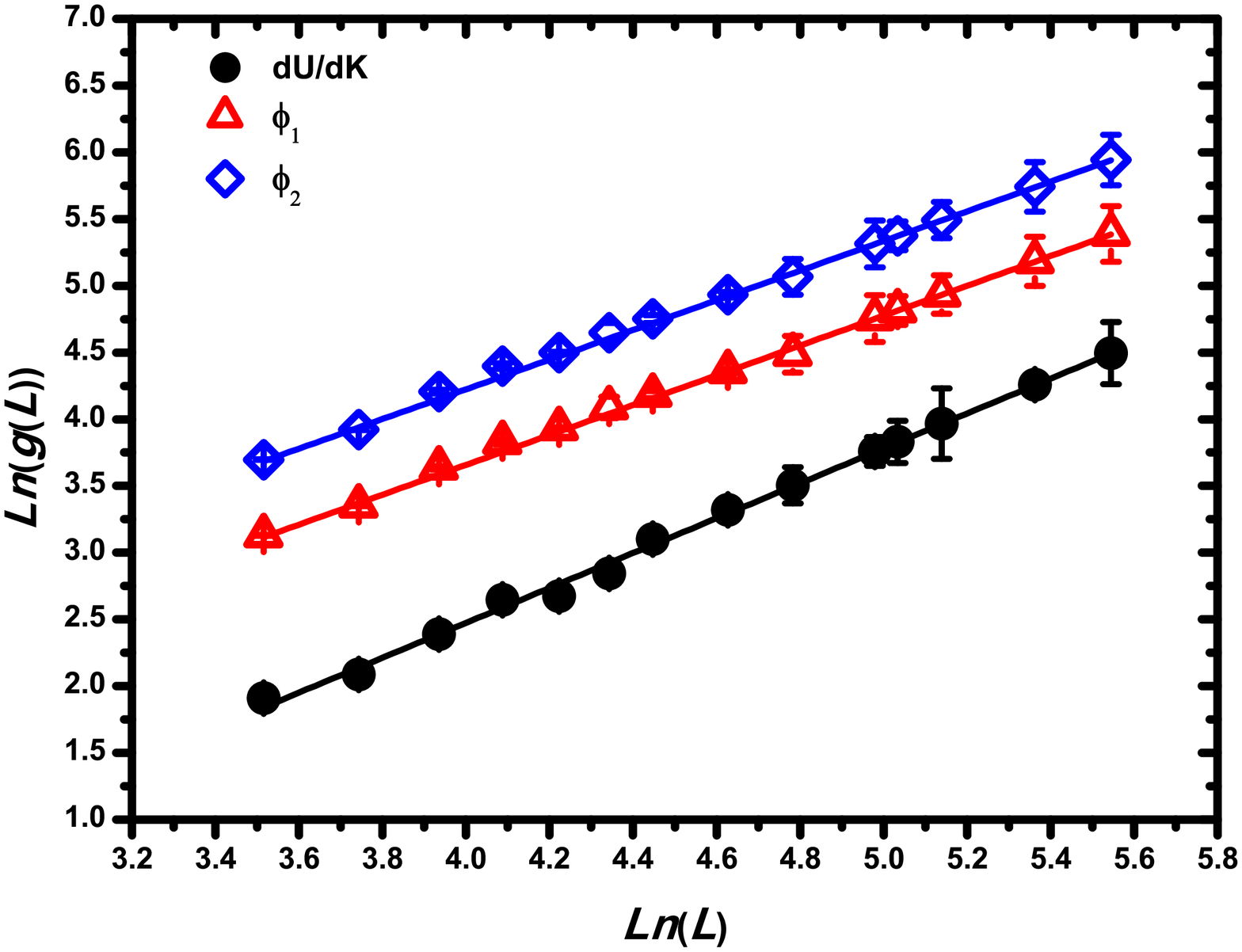}
\caption{Log-log plot of the size dependence of the maximum values of the thermodynamic derivatives $g(L) \equiv dU/dK$ (filled black circle), $\phi_1$ (red triangle) and $\phi_2$ (blue diamond) for the $q=3$ on the QDL. The simulation is performed at $k_{B}T_{0}/J=1.033$.}\label{fig:4}
\end{minipage}\hfill
\begin{minipage}[!r]{0.49\linewidth}
\includegraphics*[scale=0.30,angle=0]{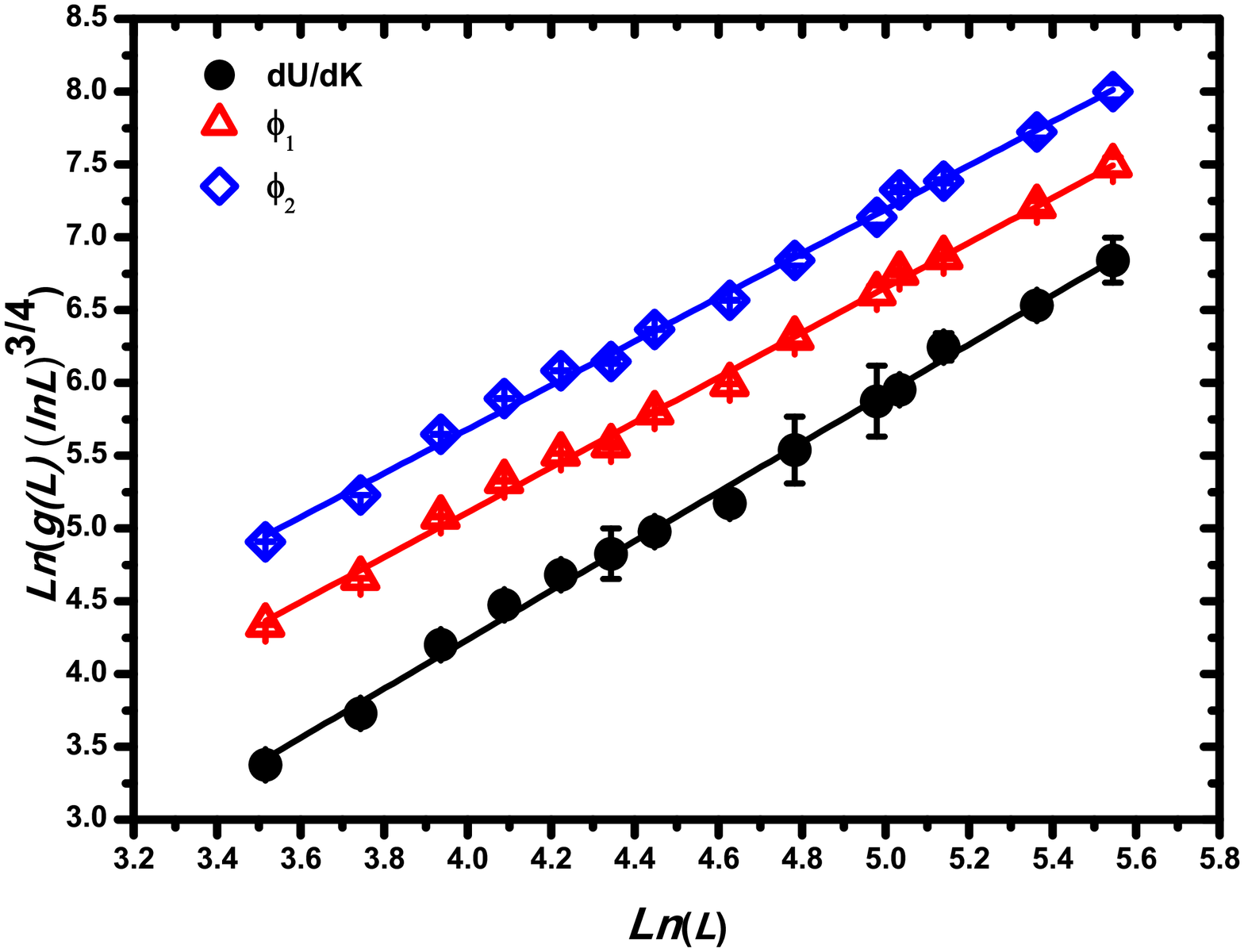}
\caption{Log-log plot of the size dependence of the maximum values of the thermodynamic derivatives $g(L)\equiv dU/dK$ (filled black circle), $\phi_1$ (red triangle) and $\phi_2$ (blue diamond) for the $q=4$ on the QDL. The simulation is performed at $k_{B}T_{0}/J=0.943$.}\label{fig:11}
\end{minipage}
\end{figure*}

\begin{table*}
\caption{\label{tab:1a} Estimates of the ratios of the leading critical exponents for $q=3$ Potts model on QDL.}
\begin{ruledtabular}
\begin{tabular}{ccccc}
 & & & &  \\
 & $\nu$ & $\alpha/\nu$ & $\beta/\nu$ & $\gamma/\nu$  \\ \\ \hline
 & & & &  \\
 $T_{0}=1.033$&$0.853\pm 0.011$&$0.439\pm 0.004$&$0.136\pm 0.003$&$1.741\pm 0.017$\\
 $T_{0}=1.035$&$0.832\pm 0.012$&$0.448\pm 0.004$&$0.130\pm 0.003$&$1.762\pm 0.015$
\end{tabular}
\end{ruledtabular}
\end{table*}

\section{Strip Projection Method and Periodic Boundary Conditions\label{sec:proc}}

The strip projection method is a powerful technique for constructing periodic and non-periodic lattices. The methodology can be summarized as follows. First, starting from a regular lattice $\mathbb{Z}^n \in \mathbb{R}^n$ whose unit cell, $\phi$, is spanned by the $n$ vectors $\{\vec a_{1}, \ldots , \vec a_n \} $, we can resolve $\mathbb{R}^n$ into two mutually orthogonal subspaces, namely, $\varepsilon^\|$ and $\varepsilon^\bot$, of dimensions $p$ and $n-p$, respectively, i.e., $\mathbb{R}^n=\varepsilon^\|\oplus\varepsilon^\bot$. Second, we define a ``strip'' $s\in\mathbb{R}^n$ as a set of all the points whose positions are obtained by adding any vector in $\varepsilon^\|$ to any vector in $\phi$, i.e., $s=\varepsilon^\|+\phi$. The required lattice, $L^\|$, is the projection in $\varepsilon^\|$ of all the points in $\mathbb{Z}^n$ that are included in the strip, i.e., $L^\|=\pi^\|(\mathbb{Z}^n\cap s)$. The requirement that any point $\vec{x} \in \mathbb{Z}^n$ lies in the strip is equivalent to the condition that the projection of $\vec x$ in $\varepsilon^\bot$ lies within the projection of $\phi$ in $\varepsilon^\bot$. This equivalence can be mathematically expressed as
\begin{equation}\label{eq:1}
 \vec{x}\in s\Leftrightarrow \vec{x}^\bot \in \phi^\bot,
\end{equation}
where $\vec{x}^\bot=\pi^\bot(\vec{x})$ and $\phi^\bot=\pi^\bot(\phi).$, Accordingly, the lattice can be defined as follows:
\begin{equation}\label{eq:2}
L^\|=\{\vec{x}^\||\vec{x}\in\mathbb{Z}^n,\vec{x}^\bot \in \phi^\bot\}.
\end{equation}
One way to describe the projection of the points $\vec{x}\in\mathbb{Z}^n$ given by $\vec{x}=\sum_{i=1}^{n}u_{i}\vec{a}_{i}$ (where the $u_{i}$'s are integers) onto $\varepsilon^\|$ and $\varepsilon^\bot$ is to choose an orthogonal basis $\{ \vec b_{1}, \ldots , \vec b_p \} $ in $\varepsilon^\|$ and an orthogonal basis $\{ \vec b_{p+1}, \ldots ,\vec b_n \}$ in $\varepsilon^\bot$. Together they form a new basis $\{ \vec b_{1}, \ldots ,\vec b_n \}$ of $\mathbb{R}^n$. Assuming $b_{i}=a_{i}$, the relationship between the two basis can be given by a rigid rotational operation. By defining a rotation matrix $\rho$, it is possible to determine the projection matrices using the following equations
\begin{equation}\label{eq:3}
\pi _{ij}^\parallel   = \sum\limits_{k = 1}^p {\frac{{\rho _{ki} \rho _{kj} }}
{{\sigma _k }}} ;\,\,\pi _{ij}^ \bot   = \sum\limits_{k = p + 1}^n {\frac{{\rho _{ki} \rho _{kj} }}
{{\sigma _k }}},
\end{equation}
where $\sigma_{k}=\sum_{j=1}^{n}\rho_{kj}^{2}$. The rotation matrix $\rho$ can be split into an $n \times p$ submatrix $\rho^\|$ and $n \times (n-p)$ submatrix $\rho^\bot$:
\begin{equation}\label{eq:4}
\rho=\left(
  \begin{array}{c}
    \rho^\| \\
    \rho^\bot\\
  \end{array}
\right).
\end{equation}
To generate the decagonal quasiperiodic lattice, the points in the finite region of a 5D hypercubic lattice ($\mathbb{Z}^5$) are projected onto an 2D subspace ($\varepsilon^\|$) only if these points are projected inside a rhombic icosahedron, which in this case is the ``strip''. The resulting quasiperiodic lattice is obtained through the standard rotation matrix:
\begin{equation}\label{eq:5}
\rho=\frac{1}{\sqrt{10}}\left(
  \begin{array}{ccccc}
    2 & 1/\tau & -\tau & -\tau & 1/\tau \\
    0 & \tau\lambda & \lambda & -\lambda & -\tau\lambda \\
    2 & -\tau & 1/\tau & 1/\tau & -\tau \\
    0 & 1 & -\tau & \tau & -1 \\
    1 & 1 & 1 & 1 & 1 \\
  \end{array}
\right),
\end{equation}

where $\tau=(1+\sqrt{5})$ and $\lambda=\sqrt{(3-\tau)}$.
The decagonal quasiperiodic lattice consists of two types of building blocks, usually represented by a fat rhombus with an acute angle of $2\pi/5$ and a thin rhombus with an acute angle of $\pi/5$, arranged according to specific matching rules. However, the quasiperiodic lattices are not suitable for Monte Carlo simulation with periodic boundary conditions. A more suitable approach is to construct a periodic approximation of these lattices. This can be realized by simply replacing the golden number $\tau$ in the sub-matrix $\rho^\bot$ of Eq.~(\ref{eq:5}) by a rational number $F_{i-1}/F_{i}$, where $F_{i-1}$ and $F_{i}$ are successive terms in the Fibonacci sequence
$$
1\quad 1 \quad 2 \quad 3 \quad 5 \quad 8 \quad 13  \quad \dots
$$
Fig.~\ref{fig:1} shows the periodic approximation of the QDL inside a square projection window. The periodic boundary conditions are imposed at lattice sites closer to the square projection window. For finite lattices, the number of nearest neighbors at a given site range from $3$ to $10$ with a mean coordination number equal to $z=3.98$. The mean coordination number is expected to be lower than $z=4$, given the existence of a small fraction ($\leq$ 1.0\%) of sites on the boundary with a coordination number lower than $3$ in the finite lattices is analyzed in this study.

\section{ Model and Monte Carlo Simulation \label{sec:mms}}

To study the critical behavior in the QDL, we updated our lattices using the Wolff algorithm~\cite{wolff}. For a fixed temperature, we define a Monte Carlo step (MCS) per spin by accumulating the flip times of all the spins and then dividing them by the total spin number. The Hamiltonian of the {\it q}-states ferromagnetic Potts model ($J>0$) can be written as
\begin{equation}\label{eq:7}
H =  - J\sum\limits_{ < i,j > } {\delta (\sigma _i ,} {\kern 1pt} {\kern 1pt} \sigma _j ),
\end{equation}
where $\delta$ is the Kronecker delta function, and the sum runs over all nearest neighbors of $\sigma_{i}$. We also define the order parameter $m$ as
\begin{equation}\label{eq:9}
m = \frac{1}
{{(q - 1)}}(N_{max}q L^{-2}  - 1),
\end{equation}

\begin{figure*}[t]
\begin{minipage} [!r]{0.49\linewidth}
\includegraphics*[scale=0.30,angle=0]{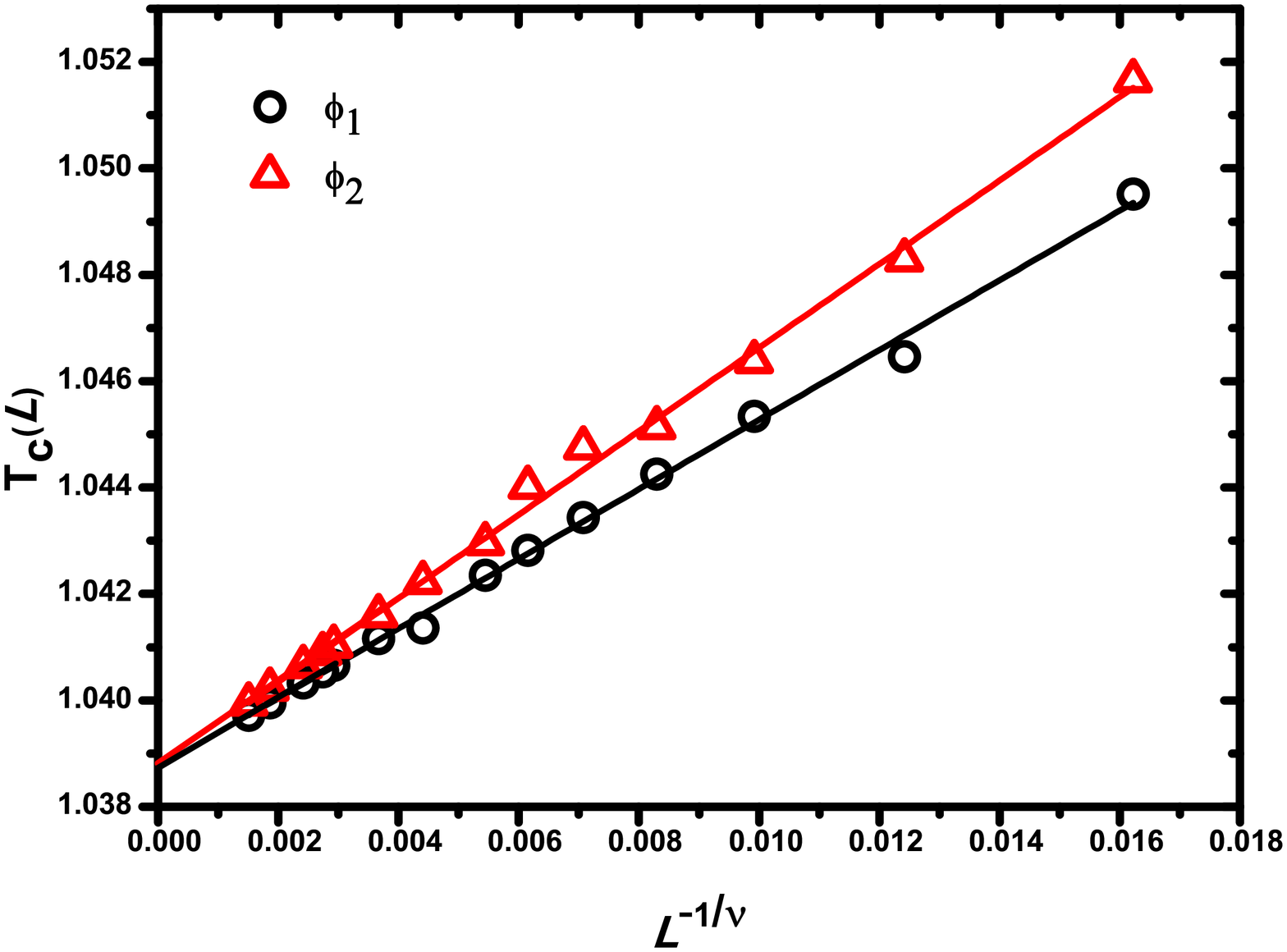}
\caption{Size dependence of the effective critical temperatures $T_c(L)$ obtained from the location of the maximum values of the thermodynamics derivatives $\phi_1$ and $\phi_2$. The curves are straight line fits to Eq.~\ref{eq:20} with $\nu=0.853$. The simulation temperature is the same as in Fig.~\ref{fig:4}.}\label{fig:5}
\end{minipage}\hfill
\begin{minipage}[!r]{0.49\linewidth}
\includegraphics*[scale=0.30,angle=0]{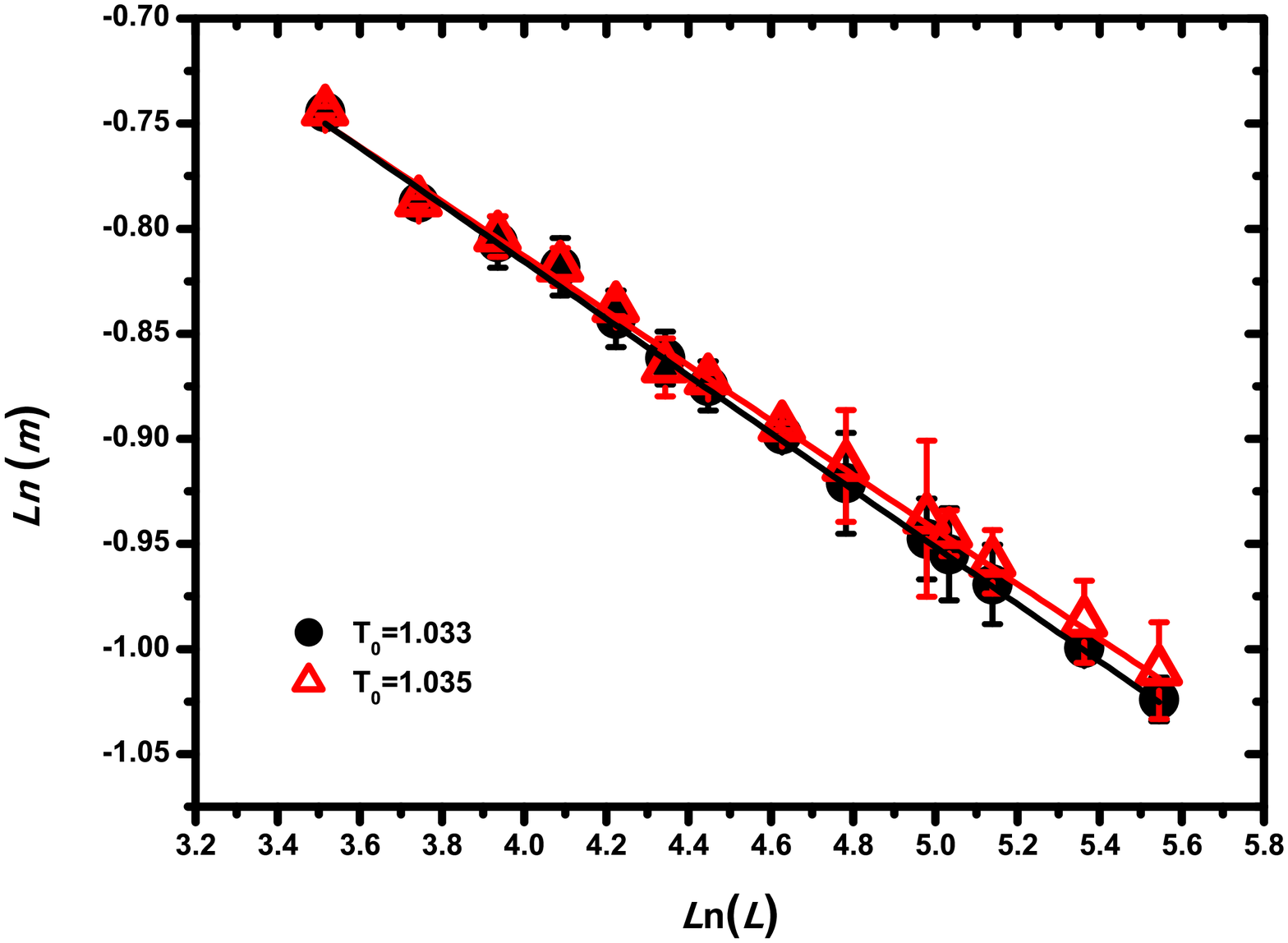}
\caption{Log-log plot of the magnetization $m$ (measured at the temperature with maximum value of $dm/dK$) versus linear size $L=\sqrt{N}$ for the $q=3$ on the QDL.}\label{fig:6}
\end{minipage}
\end{figure*}

\begin{figure*}[t]
\begin{minipage} [!r]{0.49\linewidth}
\includegraphics*[scale=0.30,angle=0]{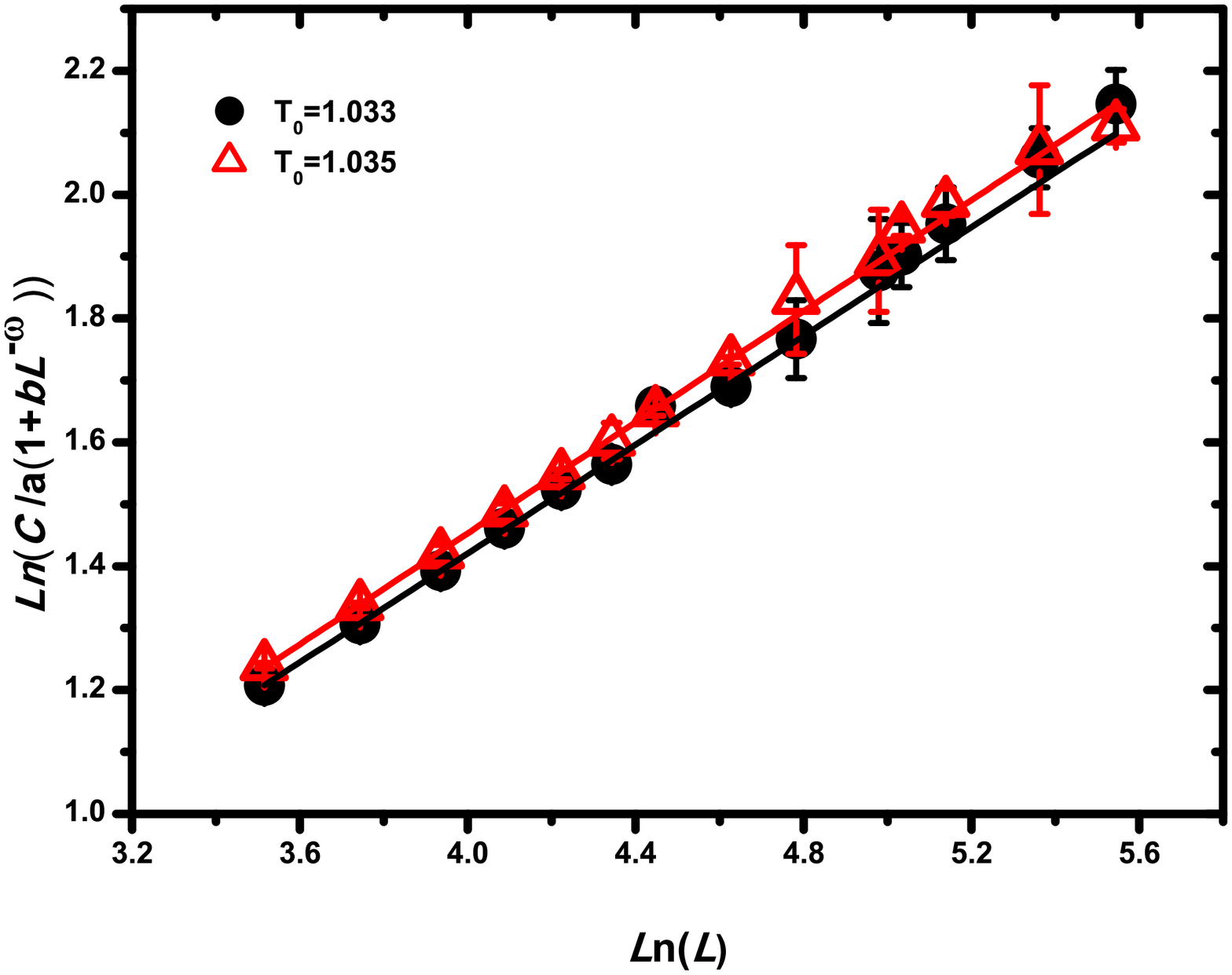}
\caption{Log-log plot of the maximum values of $C$ versus linear size $L=\sqrt{N}$ for the $q=3$ on the QDL. Data points obtained by carrying out simulations at $T_{0}=1.033$ are in filled black circles and at $T_{0}=1.035$ are in red triangles. The correlation amplitudes are $a=1.2$ and $b=2.0$ and the correction-to-scaling exponent is $\omega=1.0$. }\label{fig:7}
\end{minipage}\hfill
\begin{minipage}[!r]{0.49\linewidth}
\includegraphics*[scale=0.30,angle=0]{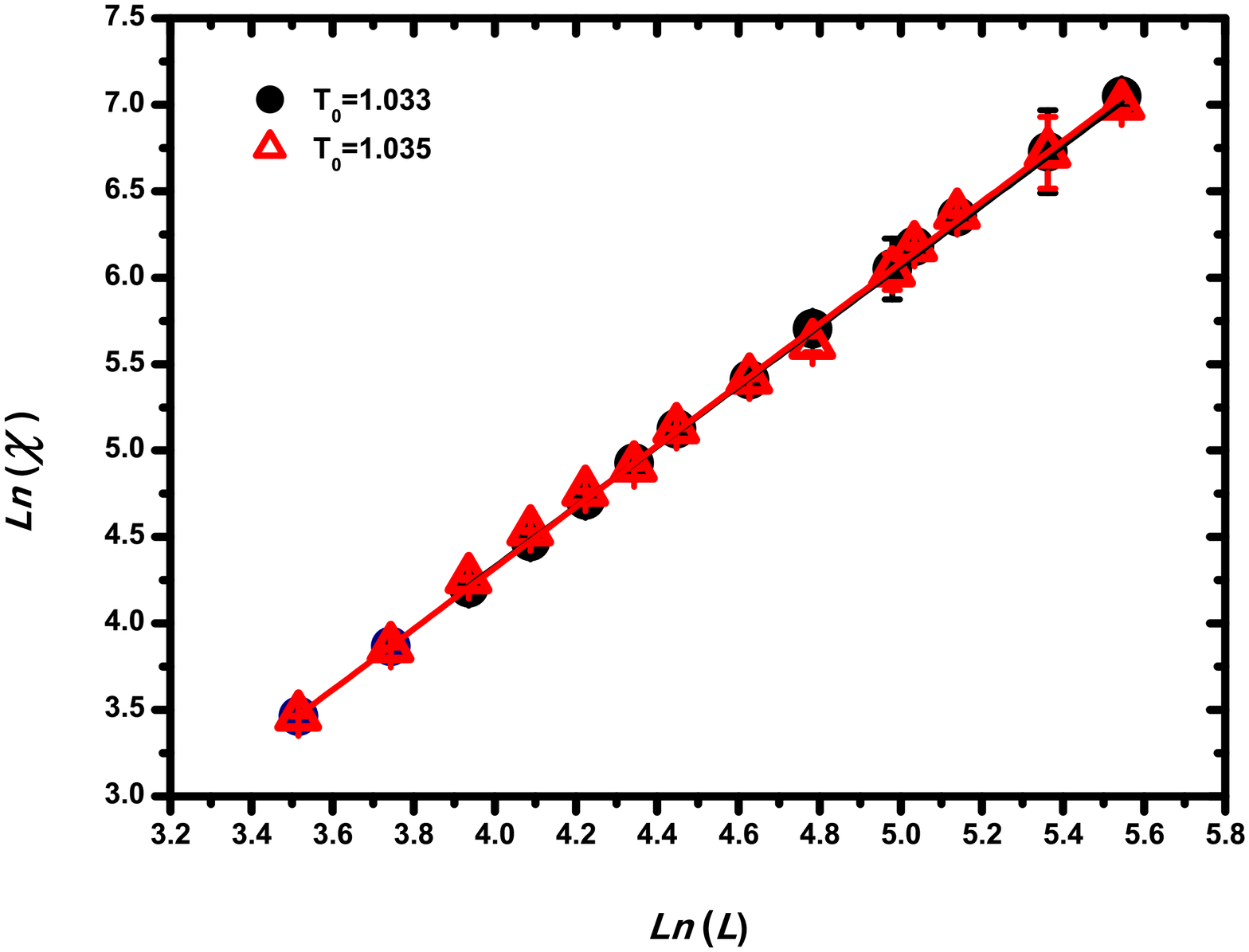}
\caption{Log-log plot of the maximum values of $\chi$ versus linear size $L=\sqrt{N}$ for the $q=3$ on the QDL.  }\label{fig:8}
\end{minipage}
\end{figure*}

\begin{table*}
\caption{\label{tab:2a} Estimates of the ratios of the leading critical exponents for $q=4$ Potts model on QDL.}
\begin{ruledtabular}
\begin{tabular}{ccccc}
 & & & &  \\
 & $\nu$ & $\alpha/\nu$ & $\beta/\nu$ & $\gamma/\nu$  \\ \\ \hline
 & & & &  \\
$T_{0}=0.940$&$0.645\pm 0.015$&$1.067\pm 0.017$&$0.118\pm 0.001$&$1.752\pm 0.010$ \\
$T_{0}=0.943$&$0.634\pm 0.010$&$1.083\pm 0.021$&$0.113\pm 0.002$&$1.795\pm 0.020$
\end{tabular}
\end{ruledtabular}
\end{table*}

where $N_{max}$ is the maximum number of spins in the same state and $L^{2}=N$ is the total number of spins.
Once the critical region is established, we apply the single histogram method~\cite{ferrenberg91,ferrenberg95} along with FSS analysis to obtain accurate estimates of the critical temperature and critical exponents.
System sizes up to $N=65391$ are used in these simulations with $1.5\times10^6$  MCS per spin performed at a single temperature $T_{0}$, where $5\times10^5$ configurations are discarded for thermalization. For each system size considered, we calculated the average of $100$ independent realizations to obtain reliable estimates of the statistical errors. The static thermodynamics quantities such as specific heat, magnetic susceptibility, logarithmic derivatives of the order parameter, and Binder's fourth-order cumulants~\cite{binder81,challa86} are then calculated inside the critical region. Depending on the analysis of the location of the maximum values of these quantities and their magnitudes, one can estimate the infinite QDL critical temperature and critical exponents, respectively. For $q=3$, we perform simulations at the temperatures $k_{B}T_{0}/J=1.033$ and $k_{B}T_{0}/J=1.035$. The probability distribution obtained at each $k_{B}T_{0}/J$ was reweighted from $k_{B}T/J=1.028$ to $k_{B}T/J=1.060$ for $N=1131, 1785, 2617, 3551, 4659$ and from $k_{B}T/J=1.028$ to $k_{B}T/J=1.048$ for $N=5919, 7285, 10445, 14271, 21111, 23543, 29117$ and $65391$. For $q=4$, we perform simulations at the temperatures $k_{B}T_{0}/J=0.940$ and $k_{B}T_{0}/J=0.943$. Here again, the probability distributions were reweighted from $k_{B}T/J=0.9300$ to $k_{B}T/J=0.960$ for the first five smaller lattices and from $k_{B}T/J=0.930$ to $k_{B}T/J=0.948$ for the remaining lattices. The specific heat can be calculated from the fluctuations of the $E$ measurements

\begin{figure*}[t]
\begin{minipage} [!r]{0.49\linewidth}
\includegraphics*[scale=0.30,angle=0]{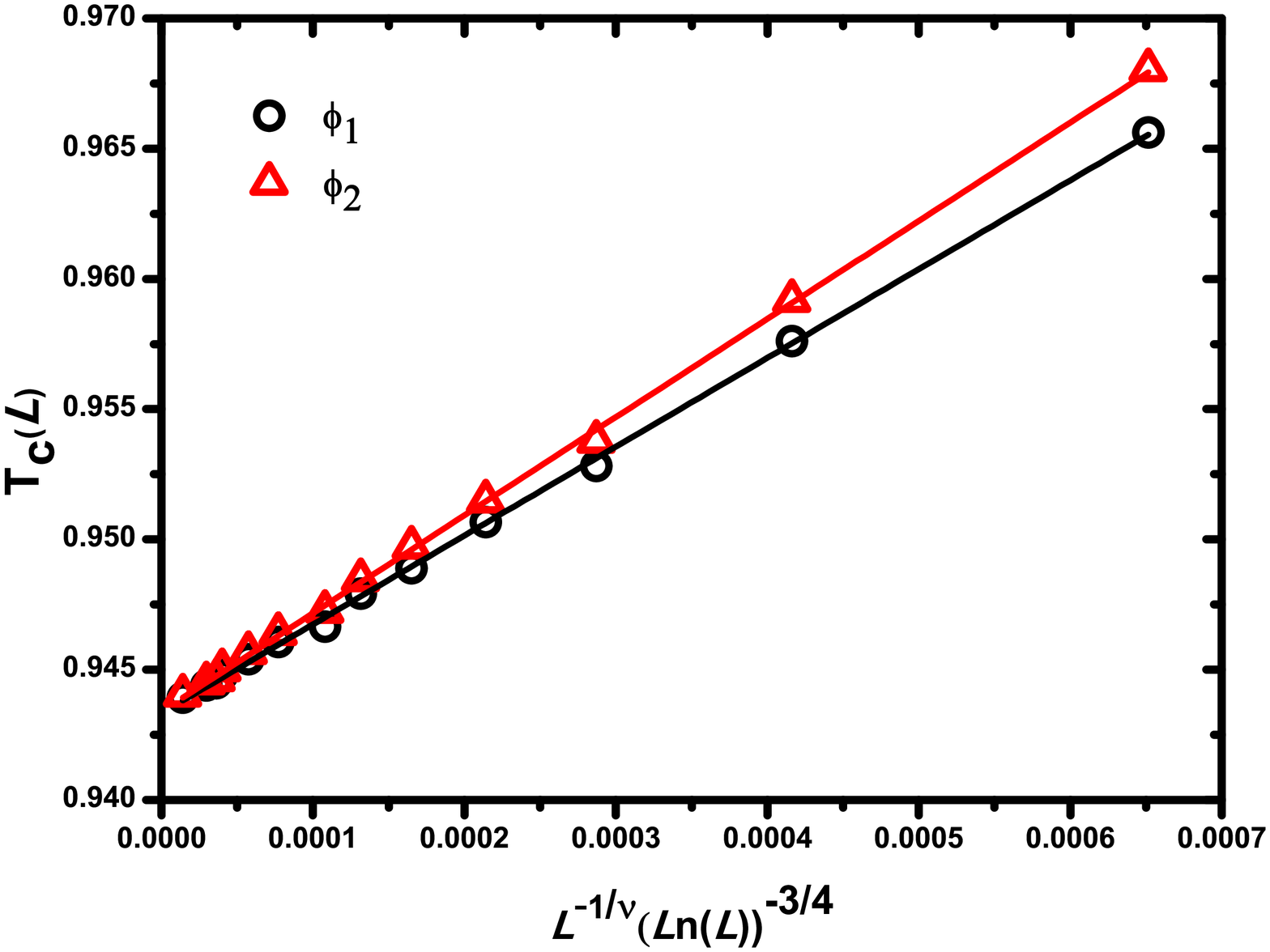}
\caption{Size dependence of the effective critical temperatures $T_c(L)$ obtained from the location of the maximum values of the thermodynamics derivatives $\phi_1$ and $\phi_2$. The curves are straight-line fits to Eq.~(\ref{eq:25}) with $\nu=0.645$.}\label{fig:12}
\end{minipage}\hfill
\begin{minipage}[!r]{0.49\linewidth}
\includegraphics*[scale=0.30,angle=0]{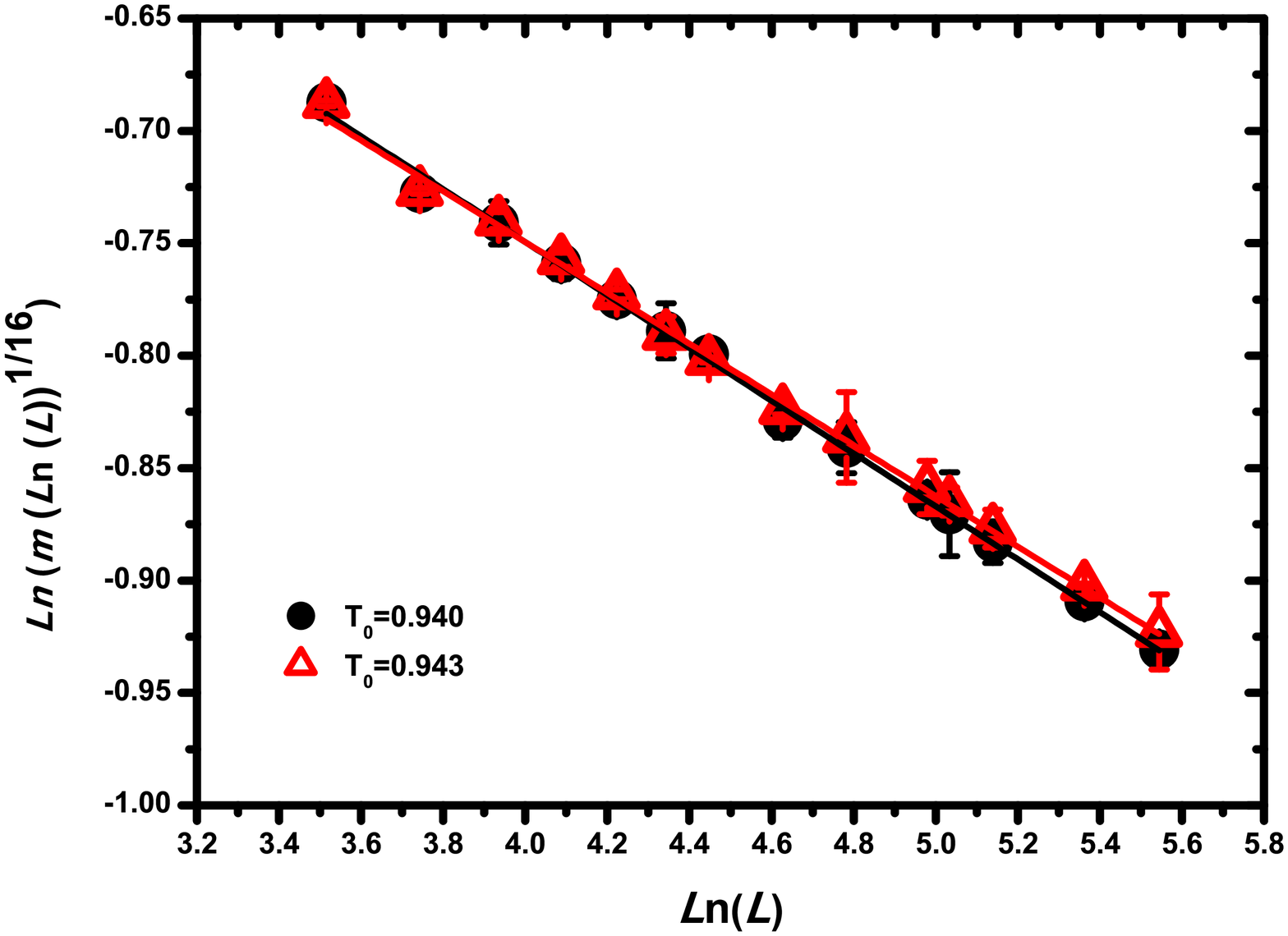}
\caption{Log-log plot of the magnetization $m$ (measured at the temperature with maximum value of $dm/dK$) versus linear size $L=\sqrt{N}$ for the $q=4$ on the QDL. }\label{fig:13}
\end{minipage}
\end{figure*}

\begin{figure*}[t]
\begin{minipage} [!r]{0.49\linewidth}
\includegraphics*[scale=0.30,angle=0]{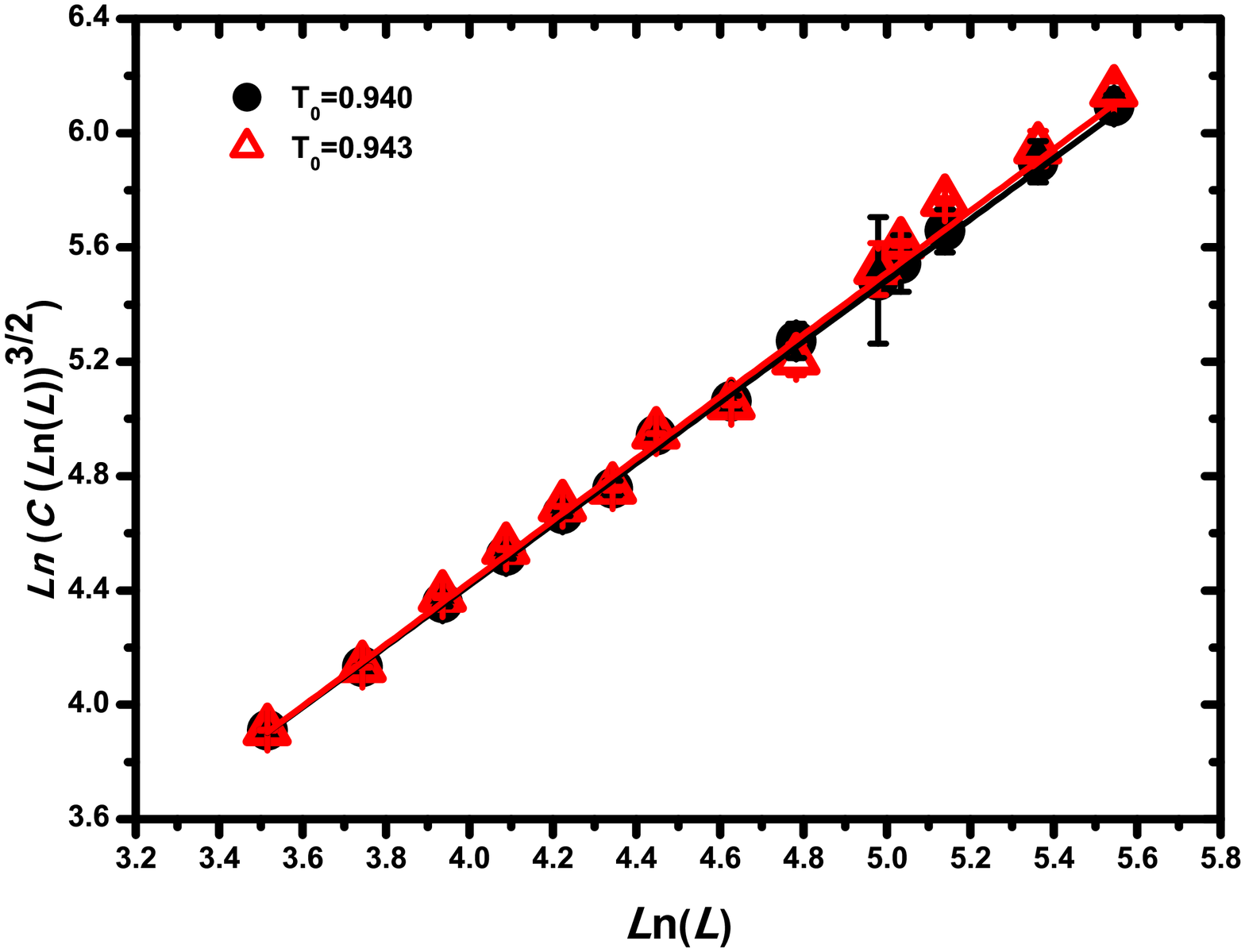}
\caption{Log-log plot of the maximum values of $C$ versus linear size $L=\sqrt{N}$ for the $q=4$ on the QDL. }\label{fig:14}
\end{minipage}\hfill
\begin{minipage}[!r]{0.49\linewidth}
\includegraphics*[scale=0.30,angle=0]{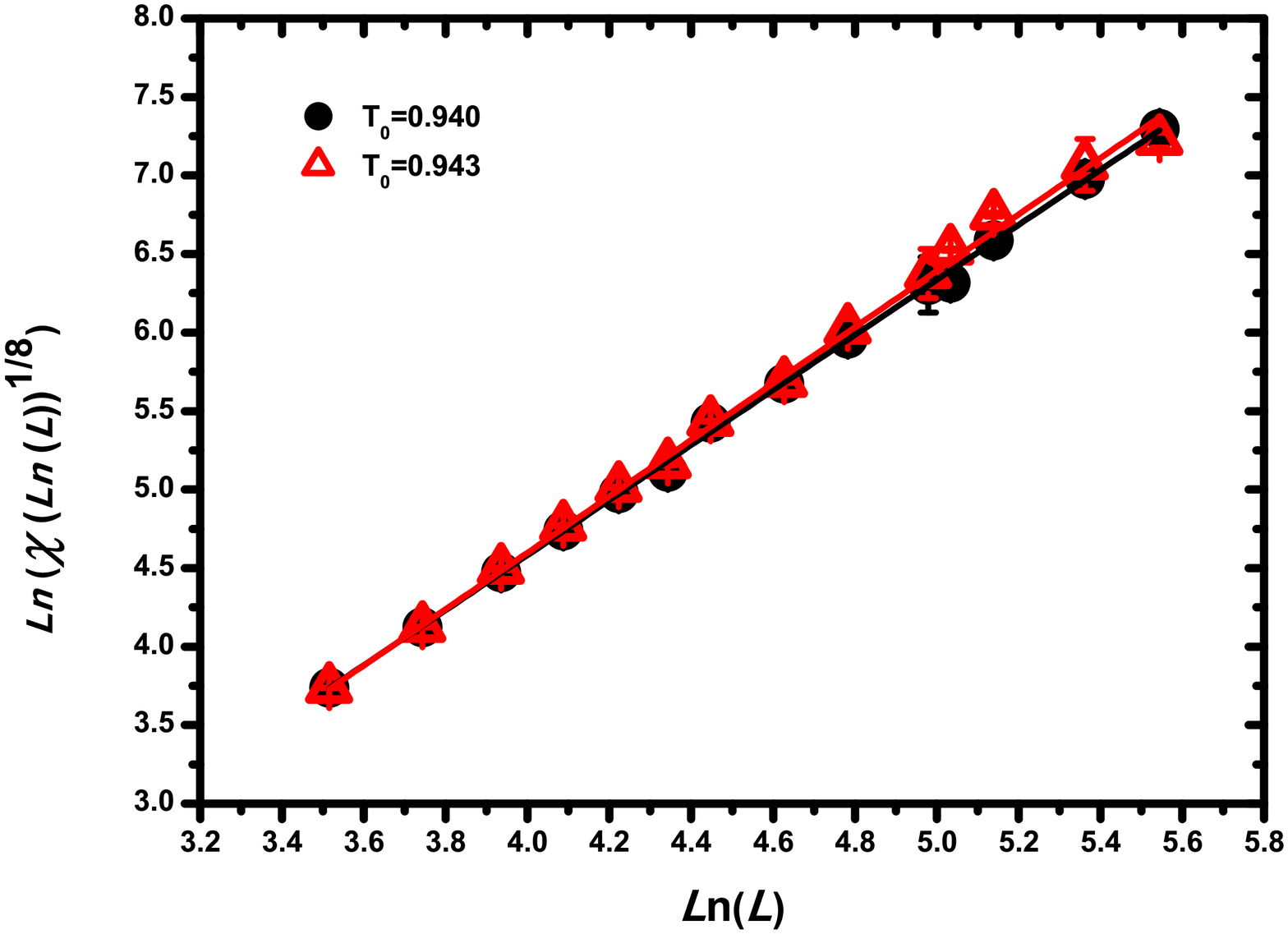}
\caption{Log-log plot of the maximum values of $\chi$ versus linear size $L=\sqrt{N}$ for the $q=4$ on the QDL.  }\label{fig:15}
\end{minipage}
\end{figure*}

\begin{equation}\label{eq:10}
C(T) = \frac{K^2}{N}( < E^2  >  -  < E > ^2 ),
\end{equation}
 Similarly, from the fluctuations of $m$, we can calculate the magnetic susceptibility
\begin{equation}\label{eq:12}
\chi (T) = KN( < m^2  >  -  < m > ^2 ),
\end{equation}
 and the fourth-order magnetization cumulant
\begin{equation}\label{eq:13}
U(T) = 1 - \frac{{ < m^4  > }}
{{3 < m^2  > ^2 }}.
\end{equation}

We can also calculate the logarithmic derivative of $n$-power of $m$, i.e,
\begin{eqnarray}
\label{eq:15}
 \nonumber  \phi_n&=& \frac{\partial}{\partial K}\,ln<m^{n}> \\
   &=& \frac{<m^{n}|\,E>}{<m^{n}>}-<E>.
\end{eqnarray}

\section{\label{sec:f} Finite-size scaling relations}

\subsection{$q=3$ Potts Model}
According to the finite-size scaling theory~\cite{fisher,fisher72}, the free energy of a system of linear dimensional $L$ is described by the scaling ansatz
\begin{equation}\label{eq:16}
f(t,h) = L^{ -d } \widetilde {f}(tL^{1/\nu } ,hL^{(\gamma  + \beta )/\nu } ),
\end{equation}
where $t=(T-T_c)/T_c$ ($T_c$ is the infinite QDL critical temperature) and $h$ is the magnetic field.
The leading critical exponents $\alpha$, $\beta$, $\gamma$ and $\nu$ define the universality class of the system.
Considering zero-field regime, the derivatives of Eq.~(\ref{eq:16}) yield important scaling equations, i.e.,
\begin{eqnarray}
  m &=& L^{ - \beta /\nu } \widetilde m(x), \label{eq:17} \\
  C &=& L^{\alpha /\nu } \widetilde C(x), \label{eq:18}\\
  \chi  &=& L^{\gamma /\nu } \widetilde \chi (x), \label{eq:19}
\end{eqnarray}
where $\widetilde m$, $\widetilde C$ and $\widetilde \chi$ are scaling functions, and $x=tL^{1/\nu}$ is the temperature scaling variable. In addition, the critical temperature scales as
\begin{equation}\label{eq:20}
    T(L)=T_c+a L^{-1/\nu},
\end{equation}
where $a$ is a constant and $T_{L}$ is the effective transition temperature for the QDL of linear size $L$.
This effective temperature can be obtained by the location of the peaks of the above quantities: $\phi_n$, $dU/dK$, $C$ and $\chi$.

\subsection{$q=4$ Potts Model}
 Due to the presence in two-dimensional $q=4$ Potts model of a marginal operator~\cite{nauenberg,cardy}, which is absent in any other two-dimensional Potts model, the leading power-law scaling behavior of this model is modified by multiplicative logarithms. So Eqs.~(\ref{eq:16}-\ref{eq:20}) must be modified to allow for these logarithmic corrections.
 The free energy scaling relation is suitably modified~\cite{aktekin,kenna} by
\begin{equation}\label{eq:21}
f(t,h) = L^{-d } \widetilde {f}(tL^{1/\nu }(lnL)^{y_t}, hL^{(\gamma+\beta )/\nu }
(lnL)^{y_h}),
 \end{equation}
 where $y_{t}=\frac{\hat{\beta}}{\beta}$ and $y_{h}=\frac{\gamma\hat{\beta}+\beta\hat{\gamma}}{2\beta+\gamma}$. Moreover, considering zero-field regime, the derivatives of Eq.(\ref{eq:21}) yield suitable scaling equations for 4-state Potts model:
 \begin{eqnarray}
  m &=& L^{ - \beta /\nu }(lnL)^{\frac{\beta\hat{\nu}}{\nu}+\hat{\beta}} \widetilde m(x), \label{eq:22} \\
  C &=& L^{\alpha /\nu }(lnL)^{-\frac{2\hat{\nu}}{\nu}} \widetilde C(x), \label{eq:23}\\
  \chi  &=& L^{\gamma /\nu }(lnL)^{-{\frac{\gamma\hat{\nu}}{\nu}+\hat{\gamma}}} \widetilde \chi (x), \label{eq:24}
\end{eqnarray}
where $\widetilde m$, $\widetilde C$ and $\widetilde \chi$ are scaling functions, and $x=tL^{1/\nu}$ is the temperature scaling variable. Correspondingly, the critical temperature scales as
\begin{equation}\label{eq:25}
    T(L)=T_c+a L^{-1/\nu}(lnL)^{\frac{\hat{\nu}}{\nu}}.
\end{equation}
In the above equations, the leading critical exponents~\cite{baxter82} for the $q=4$ Potts model on 2D periodic lattices are given by

\begin{equation}\label{eq:26}
\alpha  = \frac{2}
{3},\;\;\beta  = \frac{1}
{{12}},\;\;\gamma  = \frac{7}{6},
\;\;\nu  = \frac{2}{3},
\end{equation}
and the logarithmic-correction exponents~\cite{nauenberg,salas,berche} are given by
\begin{equation}\label{eq:27}
\hat \alpha  =  - 1,
\;\;\hat \beta  =  - \frac{1}{8},
\;\;\hat \gamma  = \frac{3}{4},
\;\;\hat \nu  = \frac{1}{2}.
\end{equation}

\section{\label{sec:r} Results}

\subsection{$q=3$ Potts Model}

Taking the slope of the log-log plot of the maximum values of the quantities $g(L)\equiv \phi_1$, $\phi_2$ and $dU/dK$ versus $L$, three estimates are obtained for $1/\nu$.   Fig.~\ref{fig:4} shows the log-log plot of these quantities for data simulated at $T_{0}=1.033$. We obtained $\nu=0.765\pm 0.013$ for $dU/dK$, $\nu=0.894\pm 0.022$ for $\phi_1$, and $\nu=0.899\pm 0.021$ for $\phi_2$. By combining these results, we get $\nu=0.853\pm 0.011$. Similar analysis has been performed for $k_{B}T_0/J=1.035$, which yielded $\nu=0.736\pm 0.019$ for $dU/dK$, $\nu=0.876\pm 0.022$ for $\phi_1$ and $\nu=0.883\pm 0.023$ for $\phi_2$. The effective range of $\nu$ is found to be $0.832\pm 0.012$ by combining the above results. These estimates are in good agreement with the exact result based on the 2D periodic lattice (with $\nu=5/6$) and in reasonable agreement with the estimates that have previously been obtained for other quasiperiodic systems~\cite{ledue97,fu2006,bin2011}.  After obtaining an estimate for $\nu$, the infinite QDL critical temperature is computed by plotting the size dependence of the location of the peaks of $\phi_1$ and $\phi_2$. Fig.~\ref{fig:5} shows the finite-size scaling of the effective transition temperatures at $k_{B}T_{0}/J=1.033$.  We obtained $T_c=1.038(7)$ for $\phi_1$ and $\phi_2$. With the simulations performed at the temperature $k_{B}T_{0}/J=1.033$, and using the corresponding $\nu$ estimated at this temperature, we obtained $T_c=1.038(8)$ for $\phi_1$ and $\phi_2$.  These values of $T_c$ are higher than the exact value on the 2D periodic lattices~\cite{kihara54}, given by $k_{B}T_c/J=1/ln(1+\sqrt{3})\approx 0.995$.

Using Eqs.~(\ref{eq:17}-\ref{eq:19}) for the size dependence of the maximum values of $m$, $C$ and $\chi$, we can estimate $\beta /\nu$, $\alpha /\nu$ and $\gamma /\nu$, respectively. Fig.~\ref{fig:6} shows the log-log plot of $m$ (measured at the temperature with maximum value of $dm/dK$) versus the linear size of the system $L$. The slopes of the linear fit to the data obtained by simulating at $T_{0}=1.035$ and $T_{0}=1.033$ are $\beta /\nu=0.130 \pm 0.003$ and $\beta /\nu=0.136 \pm 0.003$, respectively. Similarly, Fig.~\ref{fig:7} shows the log-log plot of the maximum value of $C$ versus the linear size of the system. Particularly, in this plot, we have inserted correction-to-scaling terms~\cite{shchur} to improve the fit quality of the data by scaling as
 \begin{equation}
    C= aL^{\alpha/\nu}(1+bL^{-\omega}),
\end{equation}
where the proper correlation amplitudes $a=1.2$ and $b=2.0$ and the nonuniversal correction-to-scaling exponent $\omega=1.0$ are chosen in order to minimize the $\chi^2$ of the fit. The slopes of the linear fit to the data obtained by simulating at $T_{0}=1.035$ and  $T_{0}=1.033$ are $\alpha /\nu=0.448 \pm 0.004$ and $\alpha /\nu=0.439 \pm 0.004$, respectively. Similarly, Fig.~(\ref{fig:8}) shows a log-log plot of the maximum values of $\chi$ versus $L$. The estimated values for $\gamma /\nu$ at $T_{0}=1.035$ and $T_{0}=1.033$ are $\gamma /\nu=1.762 \pm 0.015$  and $\gamma /\nu=1.741 \pm 0.017$, respectively. In these figures, the error bars are purely statistical and are estimated according to 100 different trial runs for each data point.  The estimates of the ratios of the critical exponents and the average value for $\nu$ at each simulated temperature $T_{0}$ are summarized in Table~\ref{tab:1a}.
From Table~\ref{tab:1a}, by multiplying the values of the ratios of the exponents at each simulated temperature $T_{0}$ by its respective value of $\nu$, we obtain  $\alpha=0.373 \pm 0.006$, $\beta=0.108 \pm 0.003$ and $\gamma=1.466 \pm 0.026$ at $T_{0}=1.035$, and  $\alpha=0.374 \pm 0.006$, $\beta=0.116 \pm 0.006$ and $\gamma=1.485 \pm 0.024$ at $T_{0}=1.033$.
On the 2D periodic lattice, the exact values for the $q=3$ Potts model of the critical exponents are $\nu=5/6\approx 0.833$, $\beta=1/9\approx 0.111$, $\alpha=1/3 \approx 0.333$ and $\gamma=13/9 \approx 1.444$. The final estimates of the infinite QDL critical temperature and the critical exponents for the $q=3$ Potts model on QDL are summarized and compared with the exact values on the 2D periodic lattices in Table~\ref{tab:1}.

\begin{table*}
\caption{\label{tab:1} Comparison of the final estimates of the critical temperature and leading critical exponents on QDL with the corresponding exact values on 2D periodic lattices for $q=3$ and $q=4$ Potts model.}
\begin{ruledtabular}
\begin{tabular}{ccccccc}
 & & & & & & \\
 & & $T_{C}$ & $\nu$ & $\alpha$ & $\beta$ & $\gamma$  \\ \\ \hline
 & & & & & & \\
 \multirow{3}{*}{$q=3$}& $T_{0}=1.033$ &$1.038(7)$ & $0.853 \pm 0.011$ & $0.374 \pm 0.006$ & $0.116 \pm 0.006$ & $1.485 \pm 0.024$ \\
 & $T_{0}=1.035$ & $1.038(8)$ & $0.832 \pm 0.012$ & $0.373 \pm 0.006$ & $0.108 \pm 0.003$ & $1.466 \pm 0.026$ \\
 & 2D periodic lattice & $1/ln(1+\sqrt{3})\approx 0.995$ & $5/6\approx 0.833$ & $1/3 \approx 0.333$ & $1/9\approx 0.111$ & $13/9 \approx 1.444$ \\  \hline
 & & & & & & \\
 \multirow{3}{*}{$q=4$}& $T_{0}=0.940$ & $0.943(4)$ & $0.645 \pm 0.015$ & $0.688 \pm 0.019$ & $0.076 \pm 0.002$ & $1.130 \pm 0.027$  \\
 & $T_{0}=0.943$ & $0.944(9)$ & $0.634 \pm 0.010$ & $0.687 \pm 0.017$ & $0.072 \pm 0.002$ & $1.138 \pm 0.022$  \\
 & 2D periodic lattice & $1/ln(1+\sqrt{4})\approx 0.910$ & $2/3\approx 0.667$ & $2/3 \approx 0.667$ & $1/12\approx 0.083$ & $7/6 \approx 1.167$   \\
\end{tabular}
\end{ruledtabular}
\end{table*}

\subsection{$q=4$ Potts Model}

Fig.~\ref{fig:11} shows the log-log plot of the size dependence of the maximum values of the quantities $g(L)\equiv \phi_1$, $\phi_2$ and $dU/dK$. Taking the slope of the quantities shown in Fig.~\ref{fig:11}, we obtain $\nu=0.592\pm 0.020$ for $dU/dK$, $\nu=0.648\pm 0.015$ for $\phi_1$ and $\nu=0.663\pm 0.011$ for $\phi_2$. By combining these results, we find $\nu=0.634\pm 0.010$. Similarly at $k_{B}T_0/J=0.940$, we obtain $\nu=0.610\pm 0.032$ for $dU/dK$, $\nu=0.663\pm 0.023$ for $\phi_1$ and $\nu=0.662\pm 0.022$ for $\phi_2$. Taking an average of the above results, we get $\nu=0.645\pm 0.015$. As one can see, these average values for $\nu$ are in reasonable agreement with the exact result on 2D periodic lattices, and especially for $\phi_1$ and $\phi_2$, we have a very good convergence to this exact value.

Fig.~\ref{fig:12} shows the finite-size scaling of the effective transition temperatures. From Eq.~(\ref{eq:25}) and locating the peaks of the quantities $\phi_1$ and $\phi_2$, we find $T_c=0.943(4)$. Similar analysis has been done for the temperature $k_{B}T_{0}/J=0.943$ with $\nu$ estimated at this temperature, which yielded $T_c=0.944(9)$.  Following the case $q=3$, these values are also higher than the exact value on 2D periodic lattices, given by $k_{B}T_c/J=1/ln(1+\sqrt{4})\approx 0.910$.

Using Eqs.~(\ref{eq:22}-\ref{eq:24}) and taking the exact exponents from Eqs.~\ref{eq:26} and \ref{eq:27} in the logarithmic-correction terms, we can estimate $\beta /\nu$, $\alpha /\nu$ and $\gamma /\nu$. Fig.~\ref{fig:13} shows the log-log plot of $m$ (measured at the temperature with maximum value of $dm/dK$) versus linear size of the system. The slopes of the linear fit to the data obtained by simulating at $T_{0}=0.943$ and $T_{0}=0.940$ are $\beta /\nu=0.113 \pm 0.002$ and $\beta /\nu=0.118 \pm 0.001$, respectively. Fig.~\ref{fig:14} presents a log-log plot of the maximum values of $C$ versus $L$. The slopes of the linear fit to the data obtained by simulating at $T_{0}=0.943$ and  $T_{0}=0.940$ were $\alpha /\nu=1.083 \pm 0.021$ and $\alpha /\nu=1.067 \pm 0.017$, respectively. Similarly, Fig.~\ref{fig:15} shows the log-log plot of peaks of $\chi$ versus $L$. Our estimates are $\gamma /\nu=1.795 \pm 0.020$ at $T_{0}=0.943$ and $\gamma /\nu=1.752 \pm 0.010$ at $T_{0}=0.940$.  The estimates of the ratios of the critical exponents and the average value for $\nu$ at each simulated temperature $T_{0}$ are summarized in Table~\ref{tab:2a}.

From Table~\ref{tab:2a}, by multiplying the values of the ratios of exponents at each simulated temperature $T_{0}$ by its respective value of $\nu$, we obtain $\alpha=0.687 \pm 0.017$, $\beta=0.072 \pm 0.002$ and $\gamma=1.138 \pm 0.022$ at $T_{0}=0.943$, and  $\alpha=0.688 \pm 0.019$, $\beta=0.076 \pm 0.002$ and $\gamma=1.130 \pm 0.027$ at $T_{0}=0.940$.
The exact values of the critical exponents for the $q=4$ Potts model on 2D periodic lattice are $\nu=2/3\approx 0.667$, $\beta=1/12\approx 0.083$, $\alpha=2/3 \approx 0.667$ and $\gamma=7/6 \approx 1.167$. The final estimates of the infinite QDL critical temperature and critical exponents for the $q=3$ Potts model on QDL are summarized and compared with the exact values on the 2D periodic lattices in Table~\ref{tab:1}.

\section{\label{sec:c} Conclusions}

We performed Monte Carlo simulations of the $q=3,4$-Potts model on QDL to estimate the infinite critical temperature and the leading critical exponents for both $q=3$ and $q=4$ states. Our analysis reveals that for both $q=3$ and $q=4$ states, the infinite lattice critical temperature is higher than that of the square lattice, which can be attributed to the different geometric structure between the two models. For the $q=3$ Potts model, the leading critical exponents $\nu$, $\beta$ and $\gamma$ are, within the error precision, in good agreement with the corresponding values for the 2D periodic lattices, whereas for the $q=4$ Potts model, all the critical exponents are found to be very close to the exact values on the 2D periodic lattices. This provides strong evidence to support the claim that $q=3$ and $q=4$ Potts model on quasiperiodic lattices belong to the same universality class as those on 2D periodic lattices. Future work will involve numerical studies on 3D quasiperiodic lattices so that the icosahedral phase found in alloys such as $Al$-$Fe$, $Al$-$Mn$ and $Al$-$Cr$ can also be better investigated.

\begin{acknowledgments}
We wish to thank  UFERSA for computational support and Prof. George Frederick T. da Silva for interesting discussions.
\end{acknowledgments}


\begin{thebibliography}{29}
\expandafter\ifx\csname natexlab\endcsname\relax\def\natexlab#1{#1}\fi
\expandafter\ifx\csname bibnamefont\endcsname\relax
  \def\bibnamefont#1{#1}\fi
\expandafter\ifx\csname bibfnamefont\endcsname\relax
  \def\bibfnamefont#1{#1}\fi
\expandafter\ifx\csname citenamefont\endcsname\relax
  \def\citenamefont#1{#1}\fi
\expandafter\ifx\csname url\endcsname\relax
  \def\url#1{\texttt{#1}}\fi
\expandafter\ifx\csname urlprefix\endcsname\relax\def\urlprefix{URL }\fi
\providecommand{\bibinfo}[2]{#2}
\providecommand{\eprint}[2][]{\url{#2}}

\bibitem[{\citenamefont{Shechtman}(1984)}]{shechtman84}
\bibinfo{author}{\bibfnamefont{D.}~\bibnamefont{Shechtman}},
  \bibinfo{journal}{Phys.\ Rev.\ Lett.} \textbf{\bibinfo{volume}{53}},
  \bibinfo{pages}{1951} (\bibinfo{year}{1984}).

\bibitem[{\citenamefont{Levine and Steinhardt}(1986)}]{levine86}
\bibinfo{author}{\bibfnamefont{D.}~\bibnamefont{Levine}} \bibnamefont{and}
  \bibinfo{author}{\bibfnamefont{P.~J.} \bibnamefont{Steinhardt}},
  \bibinfo{journal}{Phys.\ Rev.\ B} \textbf{\bibinfo{volume}{34}},
  \bibinfo{pages}{596} (\bibinfo{year}{1986}).

\bibitem[{\citenamefont{Potts}(1952)}]{potts52}
\bibinfo{author}{\bibfnamefont{R.~B.} \bibnamefont{Potts}},
  \bibinfo{journal}{Proc.\ Cambridge Philos.\ Soc.}
  \textbf{\bibinfo{volume}{48}}, \bibinfo{pages}{106} (\bibinfo{year}{1952}).

\bibitem[{\citenamefont{Wilson and Vause}(1988)}]{wilson88}
\bibinfo{author}{\bibfnamefont{W.~G.} \bibnamefont{Wilson}} \bibnamefont{and}
  \bibinfo{author}{\bibfnamefont{C.~A.} \bibnamefont{Vause}},
  \bibinfo{journal}{Phys.\ Lett.\ A} \textbf{\bibinfo{volume}{126}},
  \bibinfo{pages}{471} (\bibinfo{year}{1988}).

\bibitem[{\citenamefont{Wilson and Vause}(1989)}]{wilson89}
\bibinfo{author}{\bibfnamefont{W.~G.} \bibnamefont{Wilson}} \bibnamefont{and}
  \bibinfo{author}{\bibfnamefont{C.~A.} \bibnamefont{Vause}},
  \bibinfo{journal}{Phys.\ Rev.\ B} \textbf{\bibinfo{volume}{39}},
  \bibinfo{pages}{4651} (\bibinfo{year}{1989}).

\bibitem[{\citenamefont{Ledue et~al.}(1997)\citenamefont{Ledue, Boutry, Landau,
  and Teillet}}]{ledue97}
\bibinfo{author}{\bibfnamefont{D.}~\bibnamefont{Ledue}},
  \bibinfo{author}{\bibfnamefont{T.}~\bibnamefont{Boutry}},
  \bibinfo{author}{\bibfnamefont{D.~P.} \bibnamefont{Landau}},
  \bibnamefont{and} \bibinfo{author}{\bibfnamefont{J.}~\bibnamefont{Teillet}},
  \bibinfo{journal}{Phys.\ Rev.\ B} \textbf{\bibinfo{volume}{56}},
  \bibinfo{pages}{10782} (\bibinfo{year}{1997}).

\bibitem[{\citenamefont{Xiong et~al.}(1999)\citenamefont{Xiong, Zhang, and
  Tian}}]{xiong99}
\bibinfo{author}{\bibfnamefont{G.}~\bibnamefont{Xiong}},
  \bibinfo{author}{\bibfnamefont{Z.~H.} \bibnamefont{Zhang}}, \bibnamefont{and}
  \bibinfo{author}{\bibfnamefont{D.~C.} \bibnamefont{Tian}},
  \bibinfo{journal}{Phys.\ A} \textbf{\bibinfo{volume}{265}},
  \bibinfo{pages}{547} (\bibinfo{year}{1999}).

\bibitem[{\citenamefont{Fu et~al.}(2006)\citenamefont{Fu, Ma, Hou, and
  Liu}}]{fu2006}
\bibinfo{author}{\bibfnamefont{X.}~\bibnamefont{Fu}},
  \bibinfo{author}{\bibfnamefont{J.}~\bibnamefont{Ma}},
  \bibinfo{author}{\bibfnamefont{Z.}~\bibnamefont{Hou}}, \bibnamefont{and}
  \bibinfo{author}{\bibfnamefont{Y.}~\bibnamefont{Liu}},
  \bibinfo{journal}{Phys.\ Lett.\ A} \textbf{\bibinfo{volume}{351}},
  \bibinfo{pages}{435} (\bibinfo{year}{2006}).

\bibitem[{\citenamefont{Bin et~al.}(2011)\citenamefont{Bin, Lin, and
  Jun}}]{bin2011}
\bibinfo{author}{\bibfnamefont{W.~Z.} \bibnamefont{Bin}},
  \bibinfo{author}{\bibfnamefont{H.~Z.} \bibnamefont{Lin}}, \bibnamefont{and}
  \bibinfo{author}{\bibfnamefont{F.~X.} \bibnamefont{Jun}},
  \bibinfo{journal}{Chin.\ Phys.\ Lett.} \textbf{\bibinfo{volume}{28}},
  \bibinfo{pages}{046102} (\bibinfo{year}{2011}).

\bibitem[{\citenamefont{Bindi et~al.}(2015)\citenamefont{Bindi, Yao, Lin,
  Hollister, Andronicos, Distler, Eddy, Kostin, Kryachko, MacPherson
  et~al.}}]{luca}
\bibinfo{author}{\bibfnamefont{L.}~\bibnamefont{Bindi}},
  \bibinfo{author}{\bibfnamefont{N.}~\bibnamefont{Yao}},
  \bibinfo{author}{\bibfnamefont{C.}~\bibnamefont{Lin}},
  \bibinfo{author}{\bibfnamefont{L.~S.} \bibnamefont{Hollister}},
  \bibinfo{author}{\bibfnamefont{C.~L.} \bibnamefont{Andronicos}},
  \bibinfo{author}{\bibfnamefont{V.~V.} \bibnamefont{Distler}},
  \bibinfo{author}{\bibfnamefont{M.~P.} \bibnamefont{Eddy}},
  \bibinfo{author}{\bibfnamefont{A.}~\bibnamefont{Kostin}},
  \bibinfo{author}{\bibfnamefont{V.}~\bibnamefont{Kryachko}},
  \bibinfo{author}{\bibfnamefont{G.~J.} \bibnamefont{MacPherson}},
  \bibnamefont{et~al.}, \emph{\bibinfo{title}{Natural quasicrystal with
  decagonal symmetry}}, \bibinfo{howpublished}{Available from
  https://www.ncbi.nlm.nih.gov/pmc/articles/PMC4357871/}
  (\bibinfo{year}{2015}), \urlprefix\url{doi:10.1038/srep09111}.

\bibitem[{\citenamefont{Duneau and Katz}(1985)}]{duneau85}
\bibinfo{author}{\bibfnamefont{M.}~\bibnamefont{Duneau}} \bibnamefont{and}
  \bibinfo{author}{\bibfnamefont{A.}~\bibnamefont{Katz}},
  \bibinfo{journal}{Phys.\ Rev.\ Lett.} \textbf{\bibinfo{volume}{54}},
  \bibinfo{pages}{2688} (\bibinfo{year}{1985}).

\bibitem[{\citenamefont{Conway and Knowles}(1986)}]{conway86}
\bibinfo{author}{\bibfnamefont{J.~H.} \bibnamefont{Conway}} \bibnamefont{and}
  \bibinfo{author}{\bibfnamefont{K.~M.} \bibnamefont{Knowles}},
  \bibinfo{journal}{J.\ Phys.} \textbf{\bibinfo{volume}{A19}},
  \bibinfo{pages}{3645} (\bibinfo{year}{1986}).

\bibitem[{\citenamefont{Vogg and Ryder}(1996)}]{vogg96}
\bibinfo{author}{\bibfnamefont{U.}~\bibnamefont{Vogg}} \bibnamefont{and}
  \bibinfo{author}{\bibfnamefont{P.~L.} \bibnamefont{Ryder}},
  \bibinfo{journal}{J.\ Non-Cryst.\ Sol.} \textbf{\bibinfo{volume}{194}},
  \bibinfo{pages}{135} (\bibinfo{year}{1996}).

\bibitem[{\citenamefont{Wolff}(1989)}]{wolff}
\bibinfo{author}{\bibfnamefont{U.}~\bibnamefont{Wolff}},
  \bibinfo{journal}{Phys.\ Rev.\ Lett.} \textbf{\bibinfo{volume}{62}},
  \bibinfo{pages}{361} (\bibinfo{year}{1989}).

\bibitem[{\citenamefont{Ferrenberg and Landau}(1991)}]{ferrenberg91}
\bibinfo{author}{\bibfnamefont{A.~M.} \bibnamefont{Ferrenberg}}
  \bibnamefont{and} \bibinfo{author}{\bibfnamefont{D.~P.}
  \bibnamefont{Landau}}, \bibinfo{journal}{Phys.\ Rev.\ B}
  \textbf{\bibinfo{volume}{44}}, \bibinfo{pages}{5081} (\bibinfo{year}{1991}).

\bibitem[{\citenamefont{Ferrenberg et~al.}(1995)\citenamefont{Ferrenberg,
  Landau, and Swendsen}}]{ferrenberg95}
\bibinfo{author}{\bibfnamefont{A.~M.} \bibnamefont{Ferrenberg}},
  \bibinfo{author}{\bibfnamefont{D.~P.} \bibnamefont{Landau}},
  \bibnamefont{and} \bibinfo{author}{\bibfnamefont{R.~H.}
  \bibnamefont{Swendsen}}, \bibinfo{journal}{Phys.\ Rev.\ E}
  \textbf{\bibinfo{volume}{51}}, \bibinfo{pages}{5092} (\bibinfo{year}{1995}).

\bibitem[{\citenamefont{Binder}(1981)}]{binder81}
\bibinfo{author}{\bibfnamefont{K.}~\bibnamefont{Binder}}, \bibinfo{journal}{Z.\
  Phys.} \textbf{\bibinfo{volume}{43}}, \bibinfo{pages}{119}
  (\bibinfo{year}{1981}).

\bibitem[{\citenamefont{Challa et~al.}(1986)\citenamefont{Challa, Landau, and
  Binder}}]{challa86}
\bibinfo{author}{\bibfnamefont{M.~S.~S.} \bibnamefont{Challa}},
  \bibinfo{author}{\bibfnamefont{D.~P.} \bibnamefont{Landau}},
  \bibnamefont{and} \bibinfo{author}{\bibfnamefont{K.}~\bibnamefont{Binder}},
  \bibinfo{journal}{Phys.\ Lett.\ B} \textbf{\bibinfo{volume}{34}},
  \bibinfo{pages}{1841} (\bibinfo{year}{1986}).

\bibitem[{\citenamefont{Fisher}(1971)}]{fisher}
\bibinfo{author}{\bibfnamefont{M.~E.} \bibnamefont{Fisher}},
  \emph{\bibinfo{title}{Critical Phenomena}} (\bibinfo{publisher}{Academic, New
  York}, \bibinfo{year}{1971}).

\bibitem[{\citenamefont{Fisher and Barber}(1972)}]{fisher72}
\bibinfo{author}{\bibfnamefont{M.~E.} \bibnamefont{Fisher}} \bibnamefont{and}
  \bibinfo{author}{\bibfnamefont{M.~N.} \bibnamefont{Barber}},
  \bibinfo{journal}{Phys.\ Rev.\ Lett.} \textbf{\bibinfo{volume}{28}},
  \bibinfo{pages}{1516} (\bibinfo{year}{1972}).

\bibitem[{\citenamefont{Nauenberg and Scalapino}(1980)}]{nauenberg}
\bibinfo{author}{\bibfnamefont{M.}~\bibnamefont{Nauenberg}} \bibnamefont{and}
  \bibinfo{author}{\bibfnamefont{D.~J.} \bibnamefont{Scalapino}},
  \bibinfo{journal}{Phys. \ Rev.\ Lett.} \textbf{\bibinfo{volume}{44}},
  \bibinfo{pages}{837} (\bibinfo{year}{1980}).

\bibitem[{\citenamefont{J.~L.~Cardy and Scalapino}(1980)}]{cardy}
\bibinfo{author}{\bibfnamefont{M.~N.} \bibnamefont{J.~L.~Cardy}}
  \bibnamefont{and} \bibinfo{author}{\bibfnamefont{D.~J.}
  \bibnamefont{Scalapino}}, \bibinfo{journal}{Phys.\ Rev. B}
  \textbf{\bibinfo{volume}{22}}, \bibinfo{pages}{2560} (\bibinfo{year}{1980}).

\bibitem[{\citenamefont{Aktekin}(2001)}]{aktekin}
\bibinfo{author}{\bibfnamefont{N.}~\bibnamefont{Aktekin}},
  \bibinfo{journal}{J.\ Stat.\ Phys.} \textbf{\bibinfo{volume}{104}},
  \bibinfo{pages}{1397} (\bibinfo{year}{2001}).

\bibitem[{\citenamefont{Kenna}(2004)}]{kenna}
\bibinfo{author}{\bibfnamefont{R.}~\bibnamefont{Kenna}},
  \bibinfo{journal}{Nucl.\ Phys.\ B} \textbf{\bibinfo{volume}{691}},
  \bibinfo{pages}{292} (\bibinfo{year}{2004}).

\bibitem[{\citenamefont{Baxter}(1982)}]{baxter82}
\bibinfo{author}{\bibfnamefont{R.~J.} \bibnamefont{Baxter}},
  \emph{\bibinfo{title}{Exact Solved Models in Statistical Mechanics}}
  (\bibinfo{publisher}{London: Academic Press Inc}, \bibinfo{year}{1982}).

\bibitem[{\citenamefont{Salas and Sokal}(1997)}]{salas}
\bibinfo{author}{\bibfnamefont{J.}~\bibnamefont{Salas}} \bibnamefont{and}
  \bibinfo{author}{\bibfnamefont{A.~D.} \bibnamefont{Sokal}},
  \bibinfo{journal}{J.\ Stat.\ Phys.} \textbf{\bibinfo{volume}{88}},
  \bibinfo{pages}{567} (\bibinfo{year}{1997}).

\bibitem[{\citenamefont{B.~Berche and Shchur}(2009)}]{berche}
\bibinfo{author}{\bibfnamefont{W.~J.} \bibnamefont{B.~Berche},
  \bibfnamefont{P.~Butera}} \bibnamefont{and}
  \bibinfo{author}{\bibfnamefont{L.~N.} \bibnamefont{Shchur}},
  \bibinfo{journal}{Comp.\ Phys.\ Comput} \textbf{\bibinfo{volume}{180}},
  \bibinfo{pages}{493} (\bibinfo{year}{2009}).

\bibitem[{\citenamefont{Kihara et~al.}(1954)\citenamefont{Kihara, Midzuno, and
  Shizume}}]{kihara54}
\bibinfo{author}{\bibfnamefont{T.}~\bibnamefont{Kihara}},
  \bibinfo{author}{\bibfnamefont{Y.}~\bibnamefont{Midzuno}}, \bibnamefont{and}
  \bibinfo{author}{\bibfnamefont{T.}~\bibnamefont{Shizume}},
  \bibinfo{journal}{J.\ Phys.\ Soc.\ Japan} \textbf{\bibinfo{volume}{9}},
  \bibinfo{pages}{681} (\bibinfo{year}{1954}).

\bibitem[{\citenamefont{L.~N.~Shchur and Butera}(2008)}]{shchur}
\bibinfo{author}{\bibfnamefont{B.~B.} \bibnamefont{L.~N.~Shchur}}
  \bibnamefont{and} \bibinfo{author}{\bibfnamefont{P.}~\bibnamefont{Butera}},
  \bibinfo{journal}{Phys.\ Rev. B} \textbf{\bibinfo{volume}{77}},
  \bibinfo{pages}{144410} (\bibinfo{year}{2008}).

\end{thebibliography}

\end{document}